\documentclass[twocolumn,showpacs,preprintnumbers,prb]{revtex4-1}

\usepackage{graphicx}
\usepackage{dcolumn}
\usepackage{bm}
\usepackage{amsmath}
\usepackage{amssymb}
\usepackage{hyperref,color,dsfont}
\renewcommand{\vec}[1]{{{\mathbf{\boldsymbol #1}}}}
\newcommand{\hn}{\hat{n}}

\begin{document}
\title{Edge  instabilities and skyrmion creation in magnetic layers}
\author{Jan M\"uller}
\email{jmueller@thp.uni-koeln.de}
\author{Achim Rosch}
\author{Markus Garst}
\affiliation{Institut f\"ur Theoretische Physik, Universit\"at zu K\"oln, D-50937 Cologne, Germany}
\date{\today}

\begin{abstract}
We study both analytically and numerically the edge of two-dimensional ferromagnets with Dzyaloshinskii-Moriya (DM) interactions, considering both chiral magnets and magnets with interface-induced DM interactions.
We show that in the field-polarized ferromagnetic phase magnon states exist which are bound to the edge, and we calculate their spectra within a continuum field theory. Upon lowering an external magnetic field,
these bound magnons condense at a finite momentum and the edge becomes locally unstable.
Micromagnetic simulations demonstrate that this edge instability triggers the creation of a helical phase which penetrates the field-polarized state within the bulk. A subsequent increase of the magnetic field allows to create skyrmions close to the edge in a controlled manner.
\end{abstract}

% Skyrmions:                        12.39.Dc
% Vortices in magnetic thin films:  75.70.Kw
% Vortex pinning (Superconductivity)74.25.Wx
% Vortex dynamics (fluid flow)      47.32.C-
% Dynamics of magnetization         75.78.-n
% magnetic properties of nanostructures 75.75.-c
% Magnetoelectronics/Spintronics: spin transport effects 75.76.+j
\maketitle

%----------------------------------------------------------------------------------------
\section{Introduction}
%----------------------------------------------------------------------------------------

The presence of the Dzyaloshinskii-Moriya (DM) interaction in ferromagnets favours a spatial twist of the magnetization leading to modulated magnetic textures like helices and skyrmion lattices, i.e., closely packed arrangements of single skyrmions. Skyrmions are spatial configurations of the magnetization with a finite topological winding number. 
They are not only found in the thermodynamic ground state but single skyrmions also arise as topologically protected  excitations of the field-polarized state. 
Skyrmion couple very efficiently to spin currents so that ultralow current densities are already sufficient to drive skyrmion configurations\cite{Jonietz2010,Schulz2012}. Both properties, their topological protection as well as their mobility, identify skyrmions as promising candidates for elementary units of future spintronic devices, see the reviews of Refs.~\onlinecite{Nagaosa2013,Fert2013,BookSeidel}.

A prerequisite for a skyrmion technology \cite{Tomasello2014} is, however, the stabilization of magnetic skyrmion configurations at ambient conditions as well as their controlled writing and deleting.
The existence of skyrmions far below room temperature has been experimentally demonstrated some years ago in the chiral magnets with space group P2$_1$3 like MnSi \cite{Muehlbauer2009,Adams2011}, FeGe \cite{Yu2011} or Cu$_2$OSeO$_3$ \cite{Adams2012,Seki2012} as well as in magnetic mono- and bilayers\cite{Heinze2011,Bergmann2014}. 
Recently, various groups have reported the observation of skyrmion configurations at ambient temperatures. 
Co-Zn-Mn alloys with the chiral space group P4$_1$32 were shown to possess the typical phase diagram of other chiral magnets but with a skyrmion lattice phase stabilized at room temperature \cite{Tokunaga2015}. 
In addition, certain magnetic multilayers comprising magnetic Co and Fe atoms also show stable skyrmion configurations at room temperatures\cite{Woo2015,Jiang2015,MoreauLuchaire2015}. 
In all these systems, the DM interaction arises due to a lack of inversion symmetry. 
The atomic crystal of chiral magnets explicitly breaks inversion symmetry so that the DM interaction is even present in bulk samples. In magnetic multilayers, on the other hand, the DM interaction is induced by the interface between a magnetic layer and a substrate containing heavy elements.

\begin{figure}[t]
  \centering
  \begin{minipage}[b]{0.47\textwidth}
    \centering
    \includegraphics[width=0.95\textwidth]{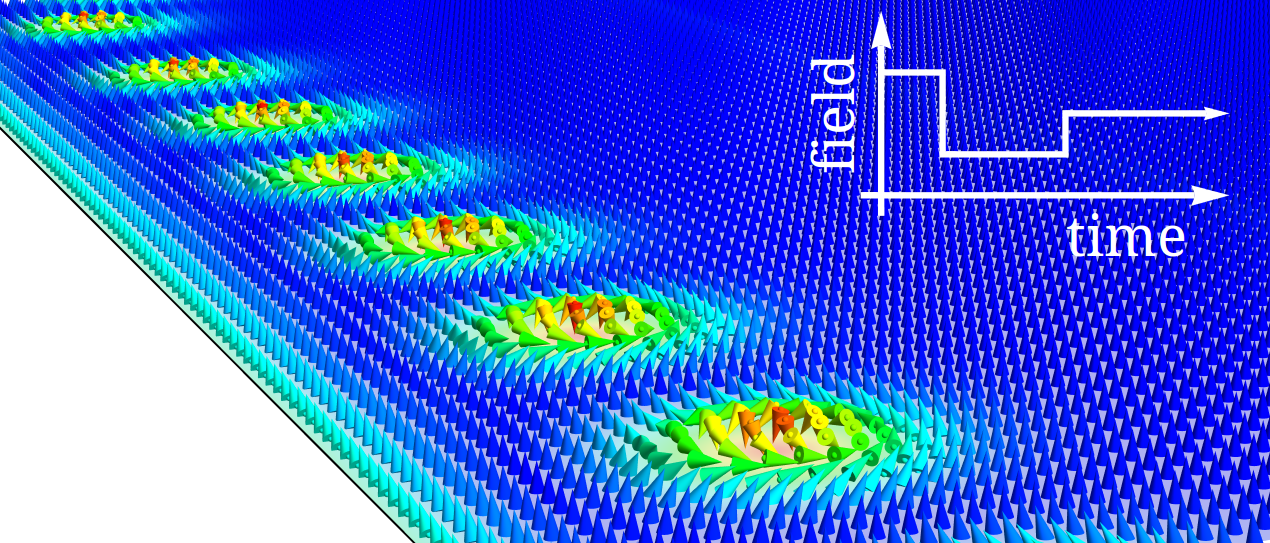}
    \caption{    Chain of skyrmions at the edge of a two-dimensional chiral magnet that were created with the help of the magnetic field protocol shown in the inset exploiting the local edge instability of the field-polarized state. These results were obtained with micromagnetic simulations of the Landau-Lifshitz-Gilbert equation. The color code reflects the $z$-component of the magnetization.}
    \label{fig1}
  \end{minipage}
\end{figure}

As skyrmion configurations are topologically protected, their creation or annihilation usually requires the climb of a topological energy barrier in some way or another. 
It has been theoretically discussed that magnetic skyrmions 
can be nucleated by 
the local injection of a spin-polarized current \cite{Sampaio2013}, local heating \cite{Koshibae2014,Koshibae2015}, local application of a magnetic \cite{Koshibae2015} or electric field \cite{Mochizuki2015}, and by thermal fluctuations \cite{Hagemeister2015}. Experimentally, 
it has been already demonstrated that skyrmions can be created in magnetic multilayers with the help of local currents from an STM tip \cite{Romming2013}. Magnetic skyrmions also materialize when driving stripe domains through a geometric constriction with the help of a spin-current \cite{Zhou2014,Jiang2015}. It was also shown that skyrmion crystals can be unwinded with the help of Bloch points\cite{Milde2013}. Such Bloch points correspond to magnetic monopoles in the emergent electrodynamics\cite{Schulz2012,Neubauer2009} that arises when itinerant electrons locally adapt their magnetic moment to the  skyrmion configurations\cite{Nagaosa2013,BookSeidel}.
%It was also shown that skyrmion crystals can be unwinded with the help of Bloch points that act as magnetic monopoles on itinerant electrons  \cite{Milde2013}.
%
Climbing the topological energy barrier and the concomitant singular Bloch point configuration of the magnetization 
 can be avoided however by feeding in skyrmions from the edge of the sample. In fact, recent experiments on FeGe nanostripes have beautifully demonstrated already the edge-mediated nucleation of skyrmions \cite{Du2015}. In the present work we provide a theoretical explanation of this phenomenon. In addition, we demonstrate that with the help of a certain magnetic field protocol skyrmions can be created in a controlled manner at the edge of a magnetic monolayer by exploiting a local edge instability of the field-polarized state, see Fig.~\ref{fig1}.

For this purpose, we investigate the edge of a single magnetic layer with DM interaction that is polarized by a perpendicular magnetic field. The boundary conditions arising from the DM interaction result in a twist of the magnetization close to the edge which was discussed independently by Wilson {\it et al.} \cite{Wilson2013} and Rohart and Thiaville {\it et al.} \cite{Rohart2013}. This surface twist was experimentally investigated in Refs.~\onlinecite{Meynell2014,Du2015}, and it also plays an important role in the stabilization of skyrmion configurations in thin films of finite thickness \cite{Rybakov2013,Rybakov2015,Rybakov2016}. 
 Here, we focus on the magnon excitations of the field-polarized layer, and we demonstrate that the surface twist result in spin-wave excitations that are bound to the edge of the magnetic layer. 
Within a continuum field-theory, we determine analytically the effective Hamiltonian of these bound magnon modes and evaluate their spectrum for various values of the effective magnetic field and the magnetic anisotropy.
For vanishing $\vec H = 0$, we reproduce the results of a recent numerical study of these edge excitations \cite{Garcia-Sanchez2014}. Moreover, for finite $\vec H$ and intermediate values of the magnetic anisotropy the edge magnons become locally unstable and condense at a finite momentum. It is this local instability that facilitates the skyrmion creation.

The outline of the paper is as follows. In sec.~\ref{sec:EdgeMagnons} we determine the magnetization at the edge of a single layer, we derive the magnon Hamiltonian and discuss its eigenmodes that are bound to the edge of the magnetic layer. 
We summerize the instabilities of the field-polarized state in sec.~\ref{sec:surface-instability} with a particular focus on the local instability triggered by the bound states that become unstable at a finite transversal momentum. 
In sec.~\ref{sec:application} we demonstrate with the help of micromagnetic simulations that this latter instability triggers the formation of a helical phase, and, in addition, how this instability can be used to create skyrmions. We conclude in sec.~\ref{sec:conc} with a short discussion.

%----------------------------------------------------------------------------------------
\section{Magnons at the edge of a chiral magnet}
\label{sec:EdgeMagnons}
%----------------------------------------------------------------------------------------

%----------------------------------------------------------------------------------------
\subsection{The Free energy functional} 
%----------------------------------------------------------------------------------------

For our study, we consider the one-dimensional edge of a two-dimensional magnetic layer. In the following, when we speak about the bulk of the system we refer in fact to the interior of this two-dimensional layer. For sufficiently low temperatures, one can neglect (thermal) amplitude fluctuations of the spins and  the local direction of the magnetization can be described by the unit vector $\hat n$. All modes discussed in the following including the boundary states at the edge  are characterized by length scales much larger than the lattice spacing, so that a description in terms of a continuum model is applicable.

The free energy functional of the magnetic layer, $F = \int d^2 \vec r \mathcal{F}$, reads 
\begin{align}
\mathcal{F} = A  (\partial_\alpha \hat{n}_i)^2 - K \hat{n}_3^2 - \mu_0 H M \hat n_3   + \mathcal{F}_{\rm DM},
\end{align}
with $\alpha = x,y$ and $i = 1,2,3$, $A$ is the stiffness, and $K$ is the magnetic anisotropy with an easy-axis and easy-plane anisotropy for $K>0$ and $K<0$, respectively. Note that for a two-dimensional layer the dipolar interactions are effectively accounted for by a renormalized anisotropy $K$\cite{SchaeferBook}.
The third term is the Zeemann energy with an external field $\vec H$ applied perpendicular to the plane, $\vec H = H \hat z$ with $H>0$, and the magnitude of the local magnetization $M$ (integrated over the $z$ direction). There a two different types of Dzyaloshinskii-Moriya (DM) interactions $\mathcal{F}_{\rm DM}$. For chiral magnets like MnSi or FeGe where the inversion symmetry is broken by the atomic crystal the DM interaction reads \cite{Bak1980}
\begin{equation}\label{Lchiral}
\mathcal{F}^{\rm chiral}_{\rm DM} = D \epsilon_{i \alpha j} \hat{n}_i \partial_\alpha \hat{n}_j ,
 \end{equation}
where $\epsilon_{i \alpha j}$ is the Levi-Civita symbol,
while for interface-induced DM interactions relevant, e.g., for Co films grown on heavy-element layers (e.g. Pt or Ir), we have \cite{Bogdanov1989,Bogdanov1994}
\begin{equation}\label{Linterface} 
\mathcal{F}^{\rm interface}_{\rm DM} = D \left( \hn_\alpha \partial_\alpha \hn_3-\hn_3 \partial_\alpha \hn_\alpha   \right) 
\end{equation}
with $\alpha = x,y$ and $i,j = 1,2,3$.
It has been pointed out in Ref.~\onlinecite{Guengoerdue2015} that for two-dimensional systems both types of DM interactions are actually equivalent because they can be transformed into each other by rotating $\hat n$ by $\pi/2$ around the $z$-axis, i.e., by $\hat n_x \to \hat n_y$, $\hat n_y \to - \hat n_x$. 
%Motivated by the experiment on FeGe films of Ref.~\onlinecite{Du2015}, 
We will use in the main part of this work the formulae appropriate for $\mathcal{F}^{\rm chiral}_{\rm DM}$ and assume $D>0$. Our results are straightforwardly generalized to interface-induced DM interactions with the help of the above symmetry.  

The dynamics of the magnetization is governed by the equation of motion
\begin{equation}\label{eom}
\partial_t \hat n =- \gamma \, \hat n \times \vec B_{\rm eff}
\end{equation}
with the gyromagnetic ratio $\gamma = g \mu_B/\hbar > 0$ and the effective magnetic field obtained by the functional derivative $\vec B_{\rm eff} = - \frac{1}{M} \delta F/\delta \hat n$ where $F = \int d^2 \vec r \mathcal{F}$.
For our micromagnetic simulations, we furthermore add Gilbert damping to these equations, see Ref.~\onlinecite{Mueller2015}.
The typical momentum, time and energy scales are given by
\begin{align} \label{LengthTime}
Q = \frac{D}{2A}, \quad  
t_{\rm DM} = \frac{\hbar M}{g \mu_B 2 A Q^2},\quad 
\varepsilon_{\rm DM} = \frac{\hbar}{t_{\rm DM}}.
\end{align}
In the main part of the paper, we will measure length, time and energy in dimensionless units by denoting the variables with a tilde, e.g.
\begin{align} \label{Units}
\tilde{x} = x Q,\quad \tilde{y} = y Q,\quad \tilde{t} = t/t_{\rm DM},\quad \tilde{\varepsilon} = \varepsilon/\varepsilon_{\rm DM}.
\end{align}
Moreover, it is convenient to introduce the dimensionless magnetic field $h$ and the dimensionless anisotropy $\kappa$, 
\begin{align}
\label{dimensionless}
h=\frac{\mu_0 H M}{2 A Q^2}, \qquad
\kappa=\frac{K}{A Q^2}.
\end{align}

As we are interested in the properties of the edge of the layered sample, we have to supplement the equations of motion
\eqref{eom} by a boundary condition. We assume the edge parallel to the $x$-axis. By varying the free energy $F=\int_{-\infty}^\infty dx \int_0^\infty  dy \mathcal F$, one obtains for chiral magnets described by Eq.~\eqref{Lchiral}  the boundary condition\cite{Meynell2014}
\begin{eqnarray}\label{boundary1}
\partial_y \hat n - Q\, \hat y \times \hat n = 0 \quad \text{at}\ y=0
\end{eqnarray}
while for interface-DM interactions \eqref{Linterface} one finds the boundary condition\cite{Rohart2013}
\begin{eqnarray}\label{boundary2}
\partial_y \hat n + Q\, \hat x \times \hat n = 0 \quad \text{at}\ y=0.
\end{eqnarray}

A subtle question is whether the continuum approach advocated above is valid for real materials. In general, the chemistry and therefore the exchange couplings, the anisotropies, and DM interactions at the edge will differ from the ones within the layer. These changes do, however, affect our results only weakly if (i) the modification of the chemical properties is limited to distances of a few lattice constants $a$ away from the edge and (ii) spin-orbit coupling effects are weak. The latter condition implies that the typical length scale on which the magnetization varies is much larger than the lattice constant $a$ and, furthermore, that anisotropy terms remain small
compared to exchange interactions even at the surface. Technically, one can check for the importance of  surface modification by adding surface, i.e., edge terms to the continuum Lagrange density, e.g., of the form $K_s \delta(y)\, \hat n_3^2$, $A_s \delta(y) \,(\partial_y \hat n_i)^2$, or $D_s \delta(y)\, \hat n_1 \partial_x \hat n_2$ with coupling constants $K_s$, $A_s$, $D_s$. The $\delta$-function takes into account that only the edge is affected. The importance of such terms can be evaluated by doing power-counting in the strength of spin orbit coupling $\lambda_{\rm SO}$ where we assume that $A \sim A_s \sim \lambda_{\rm SO}^0$, $D \sim D_s \sim \lambda_{\rm SO}$, and $K \sim K_s \sim \lambda_{\rm SO}^2$ such that the dimensionless parameters $h$ and $\kappa$ are of order 1. Under this condition all length scales are proportional to $1/\lambda_{\rm SO}$. As $\delta(y)=\lambda_{\rm SO} \delta(y \lambda_{\rm{SO}})$ all extra terms discussed above are suppressed by the small factor  $\lambda_{\rm 
SO}$ compared to the bulk contributions. This argument does nominally not hold for the surface term $\delta(y)\, \hat n \partial_y \hat n$ which by powercounting is as important as the bulk contributions  but this term is identical to zero as $\partial_y \hat n^2=0$.
For a discussion of surface terms for a three-dimensional magnet see Meynell {\it et al.}\cite{Meynell2014}.

%----------------------------------------------------------------------------------------
\subsection{The magnetization at the edge of the layer}
%----------------------------------------------------------------------------------------

We first consider the static magnetization close to the edge assuming that the magnetic layer is in a polarized state with $\hat n = \hat H = \hat z$ within the bulk of the layer. Close to the edge of the two-dimensional chiral magnet, the magnetization has to twist due to the boundary conditions (\ref{boundary1}) or (\ref{boundary2}).
An analytic solution for this twist has already been derived by Meynell {\it et al.} \cite{Meynell2014} for  
a system without magnetic anisotropy $K = 0$. Below, we generalize their result including a finite $K$. In order to simplify notations we use the rescaled coordinates of Eq.~\eqref{Units}.

The translational invariance along the edge ensures that $\hat n$ is only a function of $\tilde{y}$, i.e., the distance from the edge. We consider a chiral magnet described by Eqs. (\ref{Lchiral}) and (\ref{boundary1}) where the magnetization twists perpendicular to the propagation direction similar to a Bloch-type domain wall. Hence we choose 
\begin{equation}
\hat{n}_0^T = (\sin(\theta(\tilde y)), 0, \cos(\theta(\tilde y)))
 \label{eq:ansatz}
\end{equation}
as an ansatz for the magnetization. The equation of motion \eqref{eom} in the static limit and the boundary condition of Eq.~\eqref{boundary1}, respectively, then reduce to 
\begin{align}
&\left.\theta''(\tilde y) = h \sin(\theta(\tilde y)) + \frac{\kappa}{2} \sin(2 \theta(\tilde y)),\right.
 \label{diffE} \\
 &\left.\theta'(\tilde y) \right|_{\tilde y=0} = 1.
 \label{eq:bc_y=0}
\end{align}
Moreover, $\theta(\tilde y), \theta'(\tilde y) \to 0$ for $\tilde y \to \infty$ as we assume the magnetization $\hat n = \hat z$ to be aligned with the field within the bulk of the layer. 

\subsubsection{Kink soliton of the double sine-Gordon model}

The equation of motion Eq.~\eqref{diffE} is just the Euler-Lagrange equation of the double sine-Gordon model \cite{Condat1983,Campbell1986}. It turns out that the magnetization at the edge can be obtained by placing a kink soliton of this model close to the edge. The analytical expression for this kink soliton is known in closed form \cite{Condat1983,Campbell1986} and we shortly discuss its derivation in the following. As the derivative at the edge is positive, $\theta'(0)>0$, we will look for a kink that interpolates $\theta$ from $-2\pi$ to $0$ for increasing $\tilde y$. The first integral of the equation of motion is obtained in a standard fashion by multiplying Eq.~\eqref{diffE} with $\theta'$, integrating and using the boundary conditions within the bulk of the layer, $\theta(\infty)=\theta'(\infty)=0$,
\begin{align} \label{1Int}
\frac{\theta'^2}{2} + h \cos \theta + \frac{\kappa}{2} \cos^2\theta  = h + \frac{\kappa}{2}.
\end{align}
The kink is obtained by solving Eq.~\eqref{1Int} for $\theta'(\tilde y) > 0$. Integrating Eq.~\eqref{1Int} by separation of variables yields the kink soliton 
\begin{align} \label{EdgeMag}
\theta(\tilde{y}) = -\pi + 2 \arctan\Big(\sqrt{\frac{h}{h+\kappa}} \sinh(\sqrt{h+\kappa}(\tilde{y} - \tilde{y}_0))\Big)
\end{align}
where $\tilde{y}_0$ is an integration constant that specifies the position of the kink. For later convenience, we will also discuss the energy of the kink when it is placed deeply within the bulk, $\tilde y_0 \to \infty$. The kink energy per length $L_x$ is obtained by integrating $E_{\rm kink}/L_x = \int dy (\mathcal{F}_{\rm kink} - \mathcal{F}_{\rm FP})$ where $\mathcal{F}_{\rm kink}$ is the free energy density evaluated for the kink, and we have subtracted the energy density of the field-polarized state, $\mathcal{F}_{\rm FP}$,
\begin{align}
\frac{E_{\rm kink}}{2 A Q L_x} = -2\pi + 4\sqrt{h+\kappa} + \frac{4 h\, {\rm arctanh} \sqrt{\frac{\kappa}{h+\kappa}}}{\sqrt{\kappa}}.
\end{align}
There exists a critical magnetic field $h^{\rm cr}_{\rm kink}$ for which the kink energy vanishes,
\begin{align} \label{C-IC}
E_{\rm kink}|_{h^{\rm cr}_{\rm kink}} = 0.
\end{align}
This critical field is defined within the range $-1 < \kappa < \pi^2/4$ where it monotonically decreases from $h^{\rm cr}_{\rm kink} = 1$ to zero. For $\kappa = 0$ one recovers $h^{\rm cr}_{\rm kink} = \pi^2/16$ (Ref.~\onlinecite{Bogdanov1994}).  Whereas for larger fields the kink is an excitation of the field-polarized state, the commensurate-incommensurate transition can take place at $h^{\rm cr}_{\rm kink}$, and kinks become energetically preferred resulting in a soliton lattice. However, as the kink is a topological excitation it does not correspond to a local but rather a global instability of the field-polarized state for $h \leq h^{\rm cr}_{\rm kink}$.

\subsubsection{Edge magnetization}

The full magnetization profile is now easily obtained by placing the kink Eq.~\eqref{EdgeMag} close to the edge so that the 
boundary condition, $\theta'(0) = 1$, is fulfilled. This is achieved with the integration constant 
\begin{align} \label{Shift}
\tilde{y}_0 = -\frac{1}{\sqrt{\kappa+h}} {\rm arccosh} \Big(\frac{h+\kappa+\sqrt{(h+\kappa)^2- \kappa}}{\sqrt{h}}\Big).
\end{align}
The center of the kink is positioned outside the sample $\tilde{y}_0 < 0$. Note that there is also a solution of the boundary condition with a positive $\tilde{y}_0$ but it yields a state with larger energy. The result of Eq.~\eqref{EdgeMag} together with Eq.~\eqref{Shift} determines the magnetization close to the edge of the field-polarized magnetic layer.

\begin{figure}[t]
  \centering
  \begin{minipage}[b]{0.47\textwidth}
    \centering
    \includegraphics[width=0.95\textwidth]{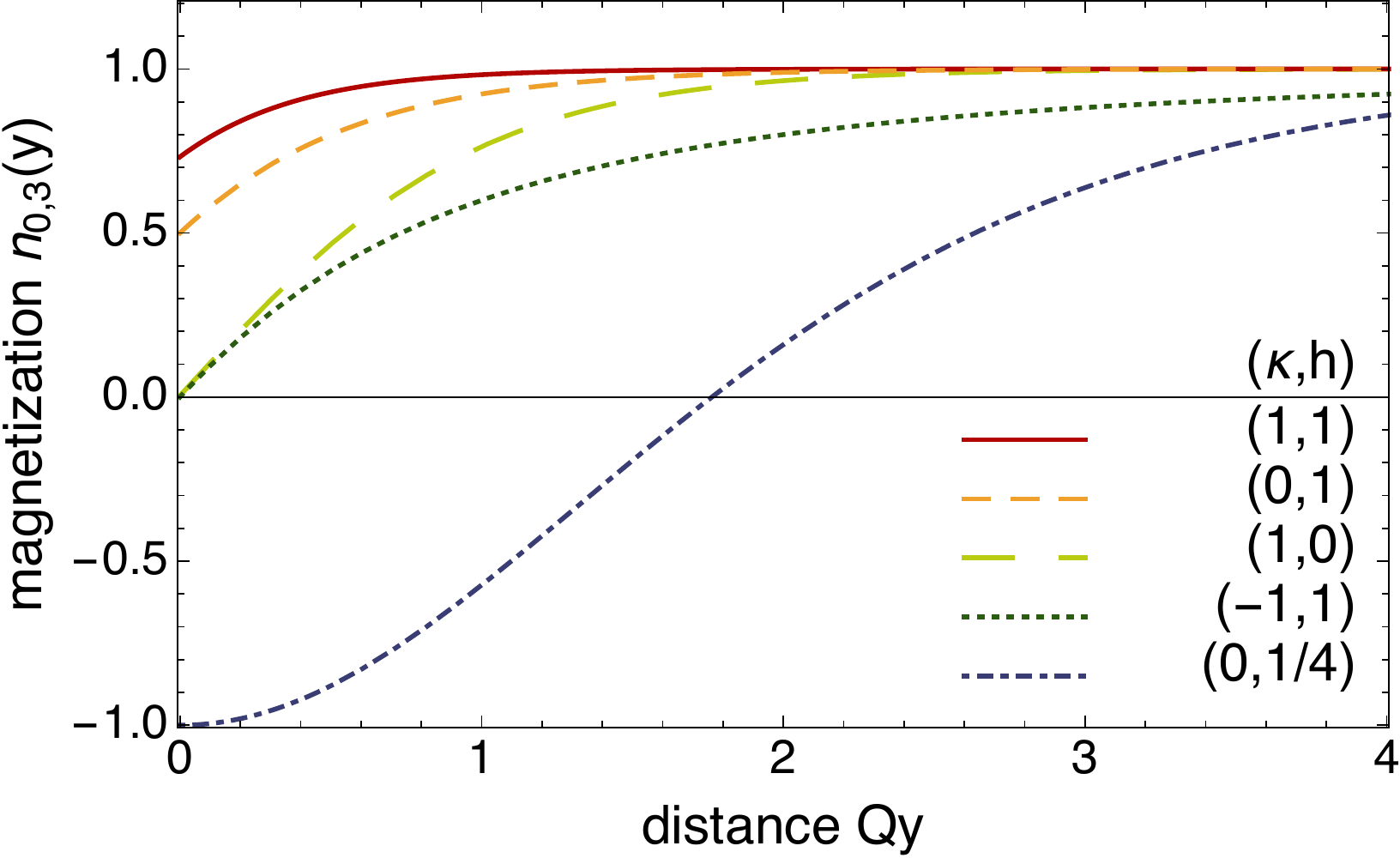}
    \caption{
Profile of the $z$-component of the normalized magnetization $\hat n_{0,3} = \cos \theta(\tilde{y})$, see Eq.~\eqref{EdgeMag}, as a function of the dimensionless distance to the edge $\tilde y = Qy$ for various values of the dimensionless field $h$ and anisotropy $\kappa$. Note that the profile for the values $(\kappa,h) = (0,1/4)$ is unstable, see discussion in sec.~\ref{sec:surface-instability}.}
    \label{fig2}
  \end{minipage}
\end{figure}

In the limit of vanishing anisotropy, $\kappa=0$, we recover the results of Ref.~\citenum{Meynell2014},
\begin{align}
\theta(\tilde y)|_{\kappa=0} &= - 4 \arctan\left(e^{- \sqrt{h}(\tilde y-\tilde y_0)} \right),
 \label{eq:instanton} \\
\tilde y_0|_{\kappa=0} &= -\frac{1}{\sqrt{h}} \ln \tan\left(\frac{1}{2} \arcsin \frac{1}{2\sqrt{h}}\right).
 \label{eq:instanton-offset}
\end{align}
We checked the result for various combinations of $h$ and $\kappa$ also numerically by simulating the damped Landau-Lifshitz-Gilbert equation and find that numerics are in excellent agreement with the analytic expression.
In Fig.~\ref{fig2} we show the analytic result for the $z$-component of the magnetization in the proximity of the edge.

%----------------------------------------------------------------------------------------
\subsection{Effective magnon Hamiltonian}
%----------------------------------------------------------------------------------------

Having established the profile of the magnetization, we now consider the fluctuations around the saddle-point solution defined by Eqs.~\eqref{eq:ansatz}, \eqref{EdgeMag} and \eqref{Shift} following the approach of Ref.~\onlinecite{Schuette2014}. We introduce a local basis 
\begin{eqnarray}
\hat{e}^T_1 =& \hat{y}^T &= (0, 1, 0) \\
\hat{e}^T_2 =& -\,\partial_\theta \hat{n}_0^T &= (-\cos\theta, 0, \sin\theta) \\
\hat{e}^T_3 =& \hat{n}_0^T &= (\sin\theta, 0, \cos\theta)
\label{eq:ansatz_basis}
\end{eqnarray}
with the angle $\theta = \theta(\tilde y)$ given by Eq.~\eqref{EdgeMag}. We use the following parametrization of the field $\hat n$, 
\begin{align}
\begin{split}
\hat{n}
&= \,\hat{e}_3 \sqrt{1-2|\psi|^2} + \hat{e}_+ \psi + \hat{e}_- \psi^\ast
\end{split}
\label{eq:ansatz_fluctuations}
\end{align}
with
$
\hat{e}_\pm = \frac{1}{\sqrt{2}} (\hat{e}_1 \pm i \hat{e}_2)
$. A change of phase of the complex magnon wavefunction $\psi = \psi(\tilde x,\tilde y,\tilde t)$ naturally encodes the precession of the magnetization. 

Expanding the Landau-Lifshitz equation \eqref{eom} up to linear order in the wavefunction $\psi$ and $\psi^\ast$, we obtain the wave equation
\begin{align}
i \tau^z \partial_{\tilde t} \vec \psi = H \vec \psi
\end{align}
for the two-component spinor $\vec{\psi} = \left(\psi \atop \psi^\ast\right)$, $\tau^i$ are Pauli-matrices, and $H$ is an effective Bogoliubov-deGennes Hamiltonian. 

It can be split into the bulk contribution $H_0$ and the edge potential $V = V(\tilde{y})$ that vanishes $V\to0$ for $\tilde y\to\infty$,
\begin{equation}
H = H_0 + V.
\label{eq:H}
\end{equation}
After performing a Fourier transformation of the spinor wavefunction, $\vec{\psi}({\tilde q_x}, \tilde y,\tilde t) = \int d\tilde x e^{-i \tilde q_x \tilde x} \vec{\psi}({\tilde x}, \tilde y,\tilde t)$, for the $\tilde x$ direction along which the problem is translationally invariant, we obtain 
\begin{equation} 
H_0 = 
-\partial^2_{\tilde y} + \tilde q_x^2 + \Delta_b,
\label{eq:H_0}
\end{equation}
where $\tilde q_x$ is the dimensionless wavevector along $\tilde x$. The magnon gap within the bulk of the layer is given by 
\begin{align} \label{bulkgap}
\Delta_b = h+\kappa.
\end{align}
For $\Delta_b < 0$, the field-polarized state within the bulk is locally unstable with respect to a tilt of the magnetization away from the field axis.
The potential reads
\begin{align}
\label{eq:V}
V(\tilde q_x,\tilde y)
&=  \mathds{1} \Big(- \theta'^2(\tilde y) +  \,\theta'(\tilde y) - \kappa \sin^2(\theta(\tilde y))\Big) 
\\\nonumber&
+ \tau^z 2 \tilde q_x  \sin(\theta(\tilde y)) 
\\\nonumber&
+ \tau^x \left(-\frac{1}{2} \theta'^2(\tilde y) +  \theta'(\tilde y) + \frac{\kappa}{2} \sin^2(\theta(\tilde y))\right) 
\end{align}
with the angle $\theta(\tilde y)$ given in Eq.~\eqref{EdgeMag}. We have used the first integral \eqref{1Int} to simplify the potential and, in particular, to eliminate the explicit dependence on dimensionless magnetic field $h$.
Moreover, an explicit calculation shows that the boundary conditions for the spinor is simply given by
\begin{equation}
\left. \partial_{\tilde y} \vec{\psi} \,\right|_{\tilde y=0} = 0 \text{.}
\label{eq:boundary_psi}
\end{equation}
The bosonic Bogoliubov-de Gennes Hamiltonian $H$, Eq.~\eqref{eq:H}, possesses scattering states that extend into the bulk of the layer as well as magnon states that a bound to the edge by the potential $V$. In the following, we will concentrate on these magnon edge modes.

%----------------------------------------------------------------------------------------
\subsection{Bound magnon edge modes}
%----------------------------------------------------------------------------------------

We will look for eigenstates $\vec \phi = \vec\phi_{\tilde \varepsilon,\tilde q}(\tilde y)$ that are localized at the edge of the chiral magnet, which requires the dimensionless energy $\tilde \varepsilon$ to have values within the range, $0 \leq \tilde \varepsilon < \Delta_b + \tilde q_x^2$. The localized eigenstates obey the stationary wave equation\cite{Schuette2014}
\begin{align}
H \vec\phi = \tilde \varepsilon \tau^z \vec \phi ,
\label{eq:Schroedinger}
\end{align}
with the normalization condition 
\begin{equation}
\int^\infty_0 \vec\phi^\dag \tau^z \vec\phi \,dy = 1 \text{.}
\label{eq:normalization}
\end{equation}
From the spinor-wavefunction $\vec\phi^T=(\phi_1,\phi_2)$ then follows the corresponding time-dependent magnon wavefunction in the parametrization of Eq.~\eqref{eq:ansatz_fluctuations}, that is $\psi(t) = \phi_1 e^{-i \varepsilon t/\hbar} + \phi_2^\ast e^{i \varepsilon t/\hbar}$ with $\varepsilon = \tilde \varepsilon \varepsilon_{\rm DM}$ where $\varepsilon_{\rm DM}$ was defined in Eq.~\eqref{LengthTime}.

We numerically search for bound state solutions using the shooting method. Starting from the boundary condition at the edge $\vec \phi^T (0)= c_1 (\cos \alpha,\sin \alpha)$ and $\vec \phi'(0)=0$ where $c_1$ is an arbitrary constant that will be fixed afterwards by the normalization condition \eqref{eq:normalization}, we vary the energy $\tilde \varepsilon$ and the parameter $\alpha$ until we find a solution of Eq.~\eqref{eq:Schroedinger} that is bound to the edge so that it decays for $\tilde y \to \infty$. The asymptotic decay of the localized wavefunction directly follows from $H_0$ of Eq.~\eqref{eq:H_0},
\begin{align}
\vec \phi(\tilde y) \sim  \left(\begin{array}{c}
c_2 e^{- \sqrt{\Delta_b+\tilde q_x^2-\tilde\varepsilon} \, \tilde y } \\
c_3 e^{- \sqrt{\Delta_b+\tilde q_x^2+\tilde\varepsilon} \, \tilde y }
\end{array}
\right)
,\quad{\rm for}\quad 
\tilde y \to \infty
\end{align}
with constants $c_2$ and $c_3$. 
The problem \eqref{eq:Schroedinger} corresponds to an effective one-dimensional wave equation, and its bound states can be labelled by a discrete quantum number $n = 0, 1,2,...$ that count the nodes of the wavefunction. Moreover, the solutions for given momenta along the edge 
$q_x$ define dispersing eigenenergies $\varepsilon_n(q_x) = \varepsilon_{\rm DM} \tilde \varepsilon_n(\tilde q_x)$ with $\tilde q_x = q_x/Q$ for the magnon bound states that are discussed in the following.

%----------------------------------------------------------------------------------------
\subsubsection{Lowest-energy  magnon edge modes}

\begin{figure}[t]
  \centering
  \begin{minipage}[b]{0.47\textwidth}
    \centering
    \includegraphics[width=0.95\textwidth]{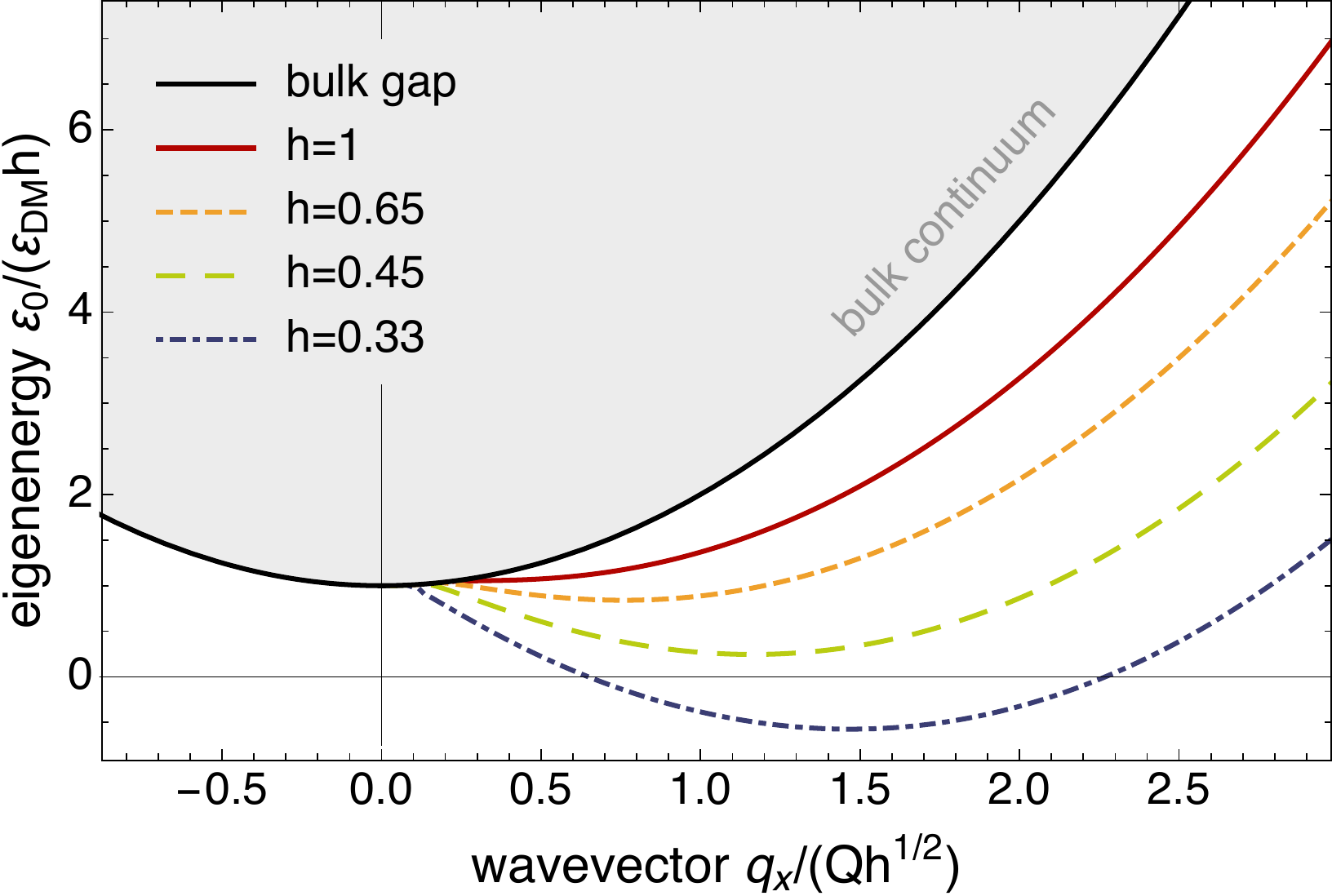}
    \caption{Dispersions $\varepsilon_0(q_x)/\varepsilon_{\rm DM} = \tilde \varepsilon_0(\tilde q_x)$ with $\tilde q_x = q_x/Q$ of the magnon edge modes with lowest energy, $n=0$, as a function of momentum $q_x$ along the edge for  fields $h=1,0.65,0.45,0.33$ and anisotropy $\kappa=0$. The grey shaded area is the bulk continuum that terminates at the solid black line.}
    \label{fig3}
  \end{minipage}
\end{figure}
\begin{figure}[t]
  \centering
  \begin{minipage}[b]{0.47\textwidth}
    \centering
    \includegraphics[width=0.95\textwidth]{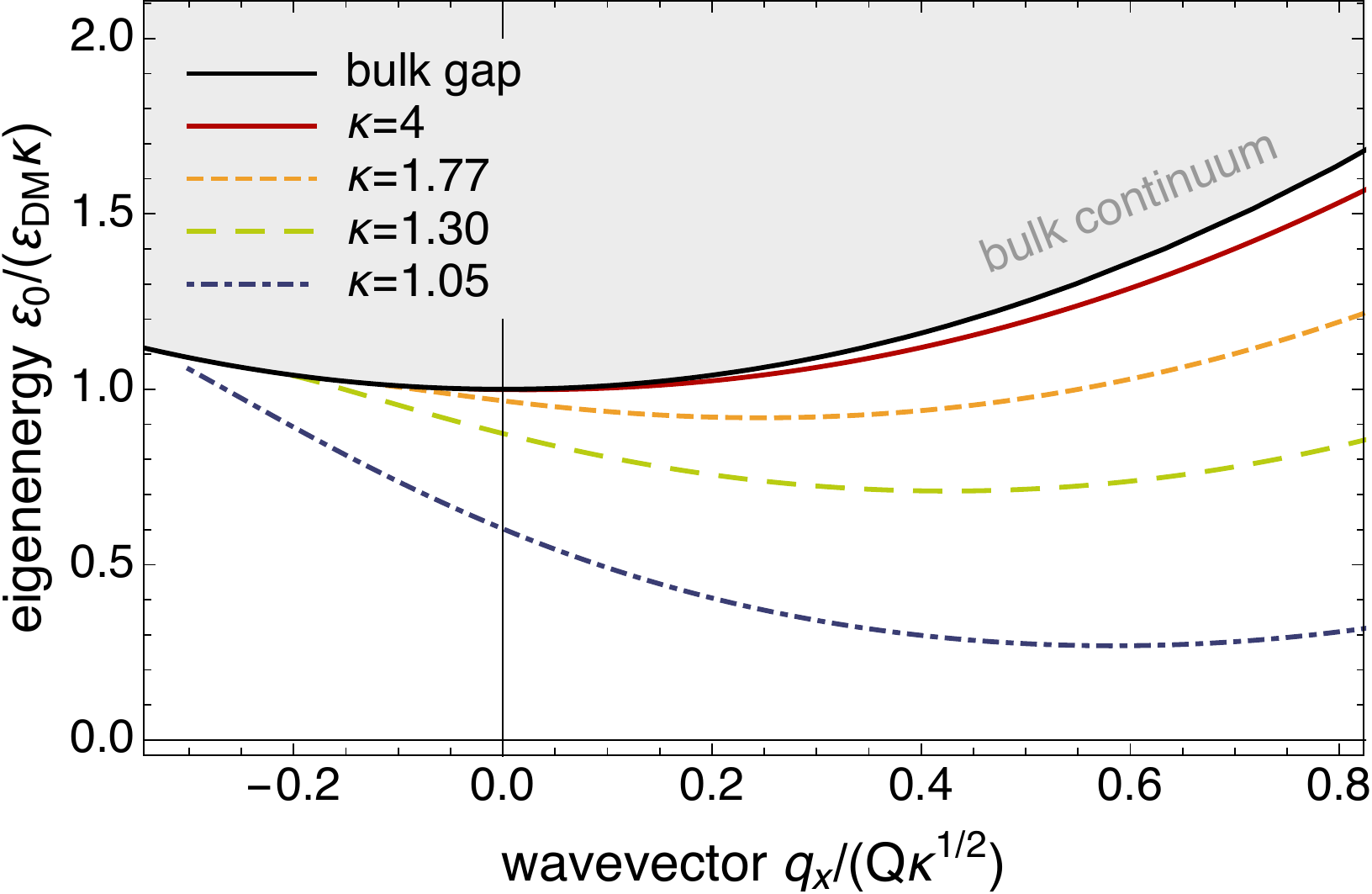}
    \caption{Dispersions $\varepsilon_0(q_x)/\varepsilon_{\rm DM} = \tilde \varepsilon_0(\tilde q_x)$ with $\tilde q_x = q_x/Q$ of the magnon edge modes with lowest energy, $n=0$, as a function of momentum $q_x$ along the edge for anisotropies $\kappa=1.05,1.30,1.77,4.00$ and field $h=0$. The grey shaded area is the bulk continuum that terminates at the solid black line.}
    \label{fig4}
  \end{minipage}
\end{figure}

The dispersion of magnon edge modes with lowest energy $\varepsilon_{0}(q_x)$, i.e., quantum number $n=0$ and momentum $q_x$ along the edge are shown for various values of the dimensionless magnetic field $h$ and $\kappa = 0$ in Fig.~\ref{fig3} as well as for various values of the dimensionless magnetic anisotropy $\kappa$ and $h =0$ in Fig.~\ref{fig4}. We have rescaled the energy by the dimensionless bulk gap $\Delta_b  =h+\kappa$ and the momenta by $\sqrt{\Delta_b}$ such that the bottom of the bulk continuum, $\Delta_b + \tilde q_x^2$, is given by a single black line for all parameters. Our results are in agreement with the numerical study by Garcia-Sanchez {\it et al.} in Ref.~\onlinecite{Garcia-Sanchez2014}.

Generally, we always find a range of momenta $q_x$ where magnon modes exist that are bound to the edge. 
Interestingly, the spectrum is chiral, i.e., it is not symmetric around $q_x=0$ as the presence of the magnetic field breaks all $q_x \to -q_x$ symmetries at the boundary. Especially for larger magnetic fields $h$ the group velocity $\partial\varepsilon_0(q_x)/\partial {q_x}$ is mostly positive, which implies that  magnons travel along the edge in a preferred direction\cite{Garcia-Sanchez2014}, for example, after they have been excited locally by a laser pulse.
Note that the modes with the reversed dispersion $\varepsilon_n(q_x) \to \varepsilon_n(-q_x)$ are found at opposite edges of the layer. Theoretically speaking, the dispersion could be reversed at the same edge by reversing the sign of the gyromagnetic ratio $\gamma \to -\gamma$, e.g., by considering holes instead of particles. However, reversing the sign of the DM interaction, $D \to -D$, modifies the edge magnetization but eventually leaves the dispersion $\varepsilon_n(q_x)$ invariant in contrast to the claim of Ref.~\onlinecite{Garcia-Sanchez2014}. This also applies for interfacial DM interaction.
While for $\kappa=0$ (Fig.~\ref{fig3})  bound states exist only for $q_x>0$, one obtains edge magnons at $q_x=0$ for $h=0$ and $\kappa> 1.005$. In the latter case an homogeneously oscillating magnetic field can be used to excite a left-moving edge mode. 
In the presence of disorder at the edge, i.e., when the  momentum $q_x$ parallel to the edge is not conserved, also right-moving modes can be excited by such an ac field.

With lower magnetic fields the minimum of the spectrum,
\begin{align} \label{EdgeGap}
\Delta_e(h,\kappa) = 
{{\rm min}} \, \tilde \varepsilon_{0}(\tilde q_x)
\end{align}
is shifted down to lower energies and to larger wave vectors. At a critical dimensionless magnetic field
$h_c = 0.4067$ at $\kappa = 0$ the minimum of the spectrum goes to zero energy, $\Delta_e = 0$, at a finite momentum $q_x$. The system therefore experiences a local instability at the edge which we discuss in further detail in Sec.~\ref{sec:surface-instability}. Similarly, we find such a local edge instability for $\kappa_c=1.005$ and $h=0$.

%----------------------------------------------------------------------------------------
\subsubsection{Higher-order magnon edge modes}

\begin{figure}[t]
  \centering
  \begin{minipage}[b]{0.47\textwidth}
    \centering
    \includegraphics[width=0.95\textwidth]{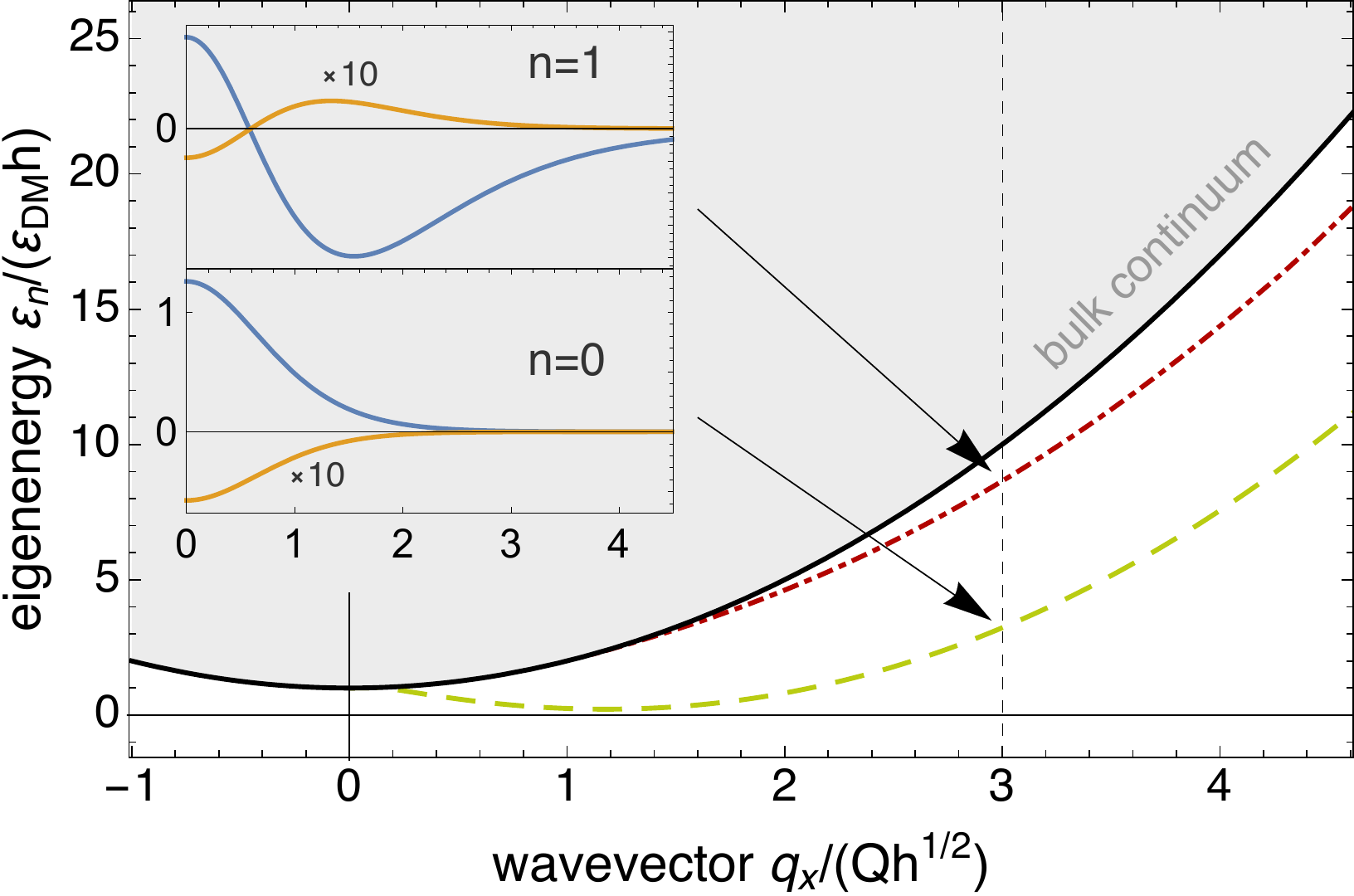}
    \caption{Dispersions $\varepsilon_n(q_x)$ of the magnon edge mode without and a single node, $n=0$ and $n =1$, respectively, for $h=0.44$ and $\kappa=0$.	      
    The inset shows the first (blue) and second (orange) component of the eigenfunction $\vec\phi$ at $q_x=3 Q$, where the latter is multiplied by a factor $10$ to make it visible. The grey shaded area is the bulk continuum that terminates at the solid black line.}
    \label{fig5}
  \end{minipage}
\end{figure}

For larger values of $q_x$, the gap between the lowest bound magnon mode $\varepsilon_0(q_x)$ and the bulk continuum increases approximately linear. This dependence can also be estimated from the potential, Eq.~\eqref{eq:V}, which grows linear in $q_x$ for large $q_x$. As the potential becomes deeper, higher-order bound states, $\varepsilon_n(q_x)$ with $n>0$, arise at the edge which are characterized by one (or more) nodes in their wave function. 
The spectrum of the bound states with zero and one node is shown in Fig.~\ref{fig5} for the magnetic field $h=0.44$.

\subsection{Phase diagram and instabilities of the metastable field-polarized state}
\label{sec:surface-instability}

In the following we discuss the phase diagram of the magnetic layer as a function of magnetic field $h$ and magnetic anisotropy $\kappa$, see Fig.~\ref{fig6}. First, we review the thermodynamically stable phases, and, afterwards, we focus on the metastable field-polarized phase and its global and local bulk and local edge instabilities.

\subsubsection{Thermodynamic phase diagram}

\begin{figure}[t]
  \centering
  \begin{minipage}[b]{0.47\textwidth}
    \centering
    \includegraphics[width=0.95\textwidth]{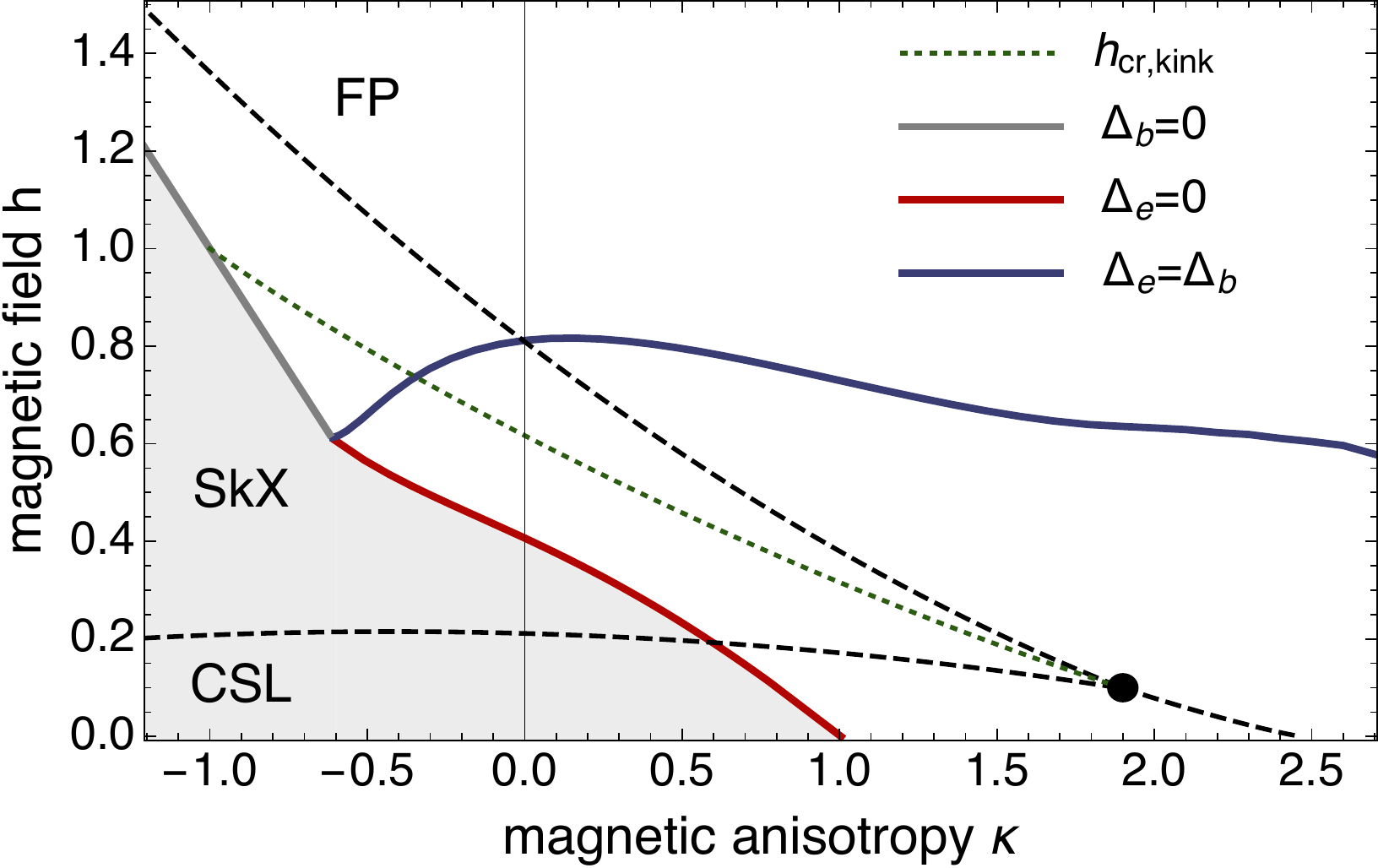}
    \caption{
    Overview of the local and global stability of the field-polarized state (FP).
    The black dashed lines are reproduced from Ref.~\onlinecite{Wilson2014} and denote thermodynamic phase transitions between the FP state, the skyrmion crystal (SkX) and the chiral soliton lattice (CSL).
    Within the white regime the FP state remains metastable, but it is locally unstable within the grey shaded regime.
   The green dotted line indicates a global instability $h^{\rm cr}_{\rm kink}$ where the energy of the kink soliton vanishes, see Eq.~\eqref{C-IC}. 
     On the grey solid line $h=-\kappa$ the magnon gap $\Delta_b$ vanishes within the bulk while the edge gap $\Delta_e$ is zero on the red solid line. 
    On the blue solid line the magnon gap within the bulk and at the edge are equal, $\Delta_e=\Delta_b$.
    }
    \label{fig6}
  \end{minipage}
\end{figure}

The bulk thermodynamic phase diagram of a magnetic layer in the presence of a perpendicular magnetic field has been studied in Refs.~\onlinecite{Wilson2014,Banerjee2014,Guengoerdue2015,Keesman2015}. For a two-dimensional chiral magnet, there exist three stable thermodynamic phases for values of $\kappa$ that are of interest here: the field-polarized state (FP), the skyrmion crystal (SkX), and the chiral soliton lattice (CSL), i.e., a helix that in general possesses higher harmonics\cite{Togawa2012}. 
The black dashed lines in Fig.~\ref{fig6} show the phase boundaries 
that were given by Wilson {\it et al.}\cite{Wilson2014}. 
(We have slightly extrapolated the phase transition lines of Ref.~\onlinecite{Wilson2014} to more negative values of $\kappa$.)
There exists a triple point (black dot) at $(\kappa,h) \approx (1.9,0.1)$ where the three phases meet\cite{Wilson2014}.

\subsubsection{Global, local bulk and local edge instabilities of the metastable field-polarized state}

Importantly, the transition between the field-polarized state (FP) and the skyrmion crystal (SkX) 
corresponds to a global instability as the two states belong to different topological sectors. If the magnetic field is reduced adiabatically at low enough temperatures to a value within the SkX regime, 
 there is the possibility that the magnetic layer remains in a metastable field-polarized state. As the field is reduced further, another global instability of the metastable FP state is encountered at $h^{\rm cr}_{\rm kink}$ within the range $-1 < \kappa < 1.9$ that is shown as the dashed green line in  Fig.~\ref{fig6}. Here the energy of a kink soliton vanishes, see Eq.~\eqref{C-IC}, triggering the formation of a chiral soliton lattice. The phase boundary between FP and CSL state within the range $1.9 < \kappa < \pi^2/4$ is also defined by $h^{\rm cr}_{\rm kink}$. 
 
The field-polarized state is stable or metastable in the whole white regime of Fig.~\ref{fig6}. It becomes locally unstable however when one of its magnon excitations reaches zero energy upon decreasing the magnetic field $h$. This instability can  occur either in the bulk or at the edge. For $\kappa < -0.61$, first the bulk  becomes locally unstable at $h=-\kappa$ (grey solid line) where the bulk gap vanishes, $\Delta_b = 0$, see Eq.~\eqref{bulkgap}. For $-0.61<\kappa < 1.005$, on the other hand, the gap of the magnon edge modes vanishes first, $\Delta_e = 0$, see Eq.~\eqref{EdgeGap}, at the red solid line. This edge instability occurs at a finite transversal momentum $q_x$, see Fig.~\ref{fig3} and Fig.~\ref{fig4}.

Finally, the blue solid line in Fig.~\ref{fig6} indicates where the bulk and edge gap have equal size. Below that line the gap $\Delta_e$ of the edge magnons is smaller than the gap $\Delta_b$ of the bulk magnons, and the spin wave excitation with lowest energies are located at the edge of the sample. 
Within this regime,  for frequencies $\Delta_e \le \hbar \tilde \omega < \Delta_b$ only edge magnons of the stable or metastable FP state are thus excited.

%----------------------------------------------------------------------------------------
\section{Edge instability and creation of skyrmions}
\label{sec:application}
%----------------------------------------------------------------------------------------

In the following we demonstrate that the local edge instability of the metastable field-polarized state at the red line in Fig.~\ref{fig6} triggers the formation of a helical state (CLS) although in some region of the phase diagram the skyrmion crystal (SkX) is in fact thermodynamically more stable. Moreover, we show that this edge instability can be exploited to create skyrmions in a controlled manner. In order to investigate the evolution of the magnetic state in the regime where the field-polarized state is locally unstable, we have performed micromagnetic simulations using the Landau-Lifshitz-Gilbert (LLG)  equation at $T=0$, for details see Ref.~\onlinecite{Mueller2015}. Typically we use $\alpha=0.1$ or $0.4$ as a damping constant in our simulations.
We simulated a two-dimensional stripe with open boundary conditions in one and periodic boundary conditions in the perpendicular direction.

\begin{figure}[t]
  \centering
  \begin{minipage}[b]{0.47\textwidth}
    \centering
    \includegraphics[width=0.95\textwidth]{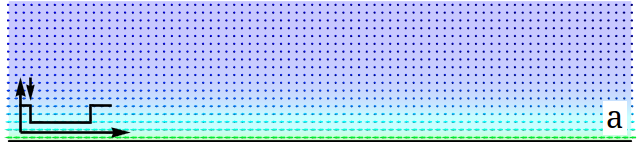}
    \includegraphics[width=0.95\textwidth]{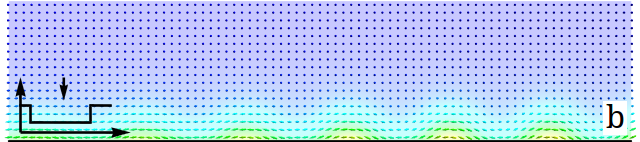}
    \includegraphics[width=0.95\textwidth]{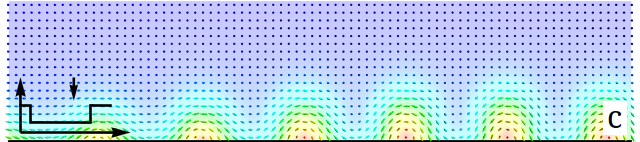}
    \includegraphics[width=0.95\textwidth]{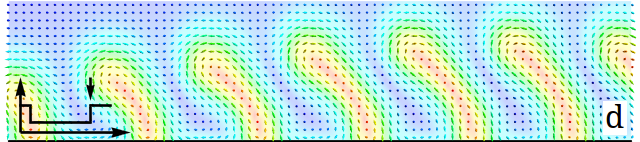}
    \includegraphics[width=0.95\textwidth]{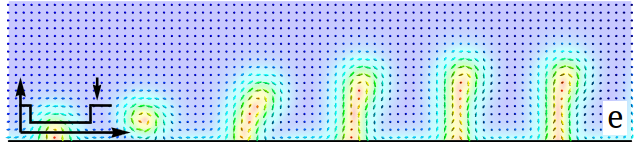}
    \includegraphics[width=0.95\textwidth]{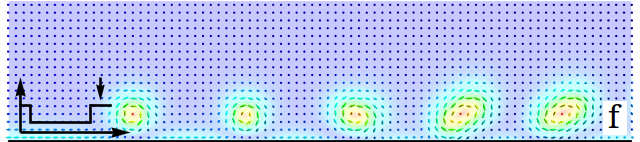}
    \includegraphics[width=0.95\textwidth]{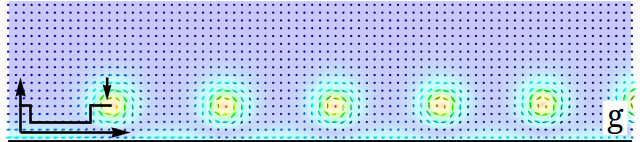}
  \end{minipage}
  \caption{Creation of a chain of skyrmions using the local edge instability of the field-polarized state. The panels show snapshots at various times obtained by LLG simulations of the field-protocol in Eq.~\eqref{protocol} as sketched in the inset with Gilbert damping $\alpha=0.4$, $\kappa=0$,  $h_i = 1$, $h_0 = 0.39$, $h_f =1$, and
$\tilde{t}_f = 40.5$.
(a)-(d) Triggered by the edge instability, a chain of merons forms that penetrates into the field-polarized state of the bulk. (e)-(g) The merons are pushed back towards the edge and their local dynamics close to the edge results in the formation of a chain of skyrmions. Steps (e)-(g) have been observed experimentally in Ref.~\onlinecite{Du2015}.
  The color code denotes the $z$-component of the magnetization.
  }
  \label{fig7}
\end{figure}

We focus on the range of magnetic anisotropies where the edge magnons locally destabilize the FP state, i.e., 
$-0.61<\kappa <1.005$
for the two-dimensional system, see Fig.~\ref{fig6}. 
%For a film of finite thickness for which the phase diagram also contains a conical phase\cite{Wilson2014,Banerjee2014}, the values of $\kappa$ would be restricted to a smaller interval.  
The following protocol for the time-dependent magnetic field allows for a controlled creation of a skyrmion chain close to the edge
\begin{align} \label{protocol}
h(t) = \left\{\begin{array}{lll}
h_i & {\rm for} & t<0 \\
h_0 & {\rm for} & 0 < t < t_{f}\\
h_f & {\rm for} &  t_{f} < t.
\end{array}
\right.
\end{align}
The initial field value $h_i$ is located above the upper black dashed line in Fig.~\ref{fig6} so that we start with a magnetization which is field-polarized. At $t=0$ the field is reduced $h_0 < h_i$ below the red line in Fig.~\ref{fig6} so that $\Delta_e < 0$. Finally, at  time $t_f$ we increase the field again to a value, $h_f > h^{\rm cr}_{\rm kink}$, that is located above the green dashed line in Fig.~\ref{fig6}. The result of the micromagnetic simulation for such a protocol at $\kappa=0$ is shown in Fig.~\ref{fig7}.

Initially, the magnetization is fully polarized except close to the lower edge of Fig.~\ref{fig7}(a) that shows the surface twist of Eq.~\eqref{EdgeMag}. As the field is lowered for $t>0$, the edge magnon becomes soft at a finite transversal momentum $q_x$ and destabilizes the magnetization whose time evolution is shown in Fig.~\ref{fig7}(b)-(d). To trigger the edge instability in the numerics, we explicitly broke translation symmetry by introducing a tiny perturbation by canting one spin at
the right-hand side of the boundary by $1\%$.
First, a periodic modulation of the edge spins grows in amplitude -- the edge magnon with negative energy becomes macroscopically occupied at finite momentum $q_x$. This state evolves smoothly into a helical state which penetrates into the field-polarized state of the bulk. The interface between the helical and polarized phase is thereby described by a chain of merons, i.e., half-skyrmions with winding number $1/2$. As a function of time, the interface moves further and further into the field-polarized state of the bulk. For this simulation the intermediate field value $h_0$ of Eq.~\eqref{protocol} was chosen to be located within the regime where the skyrmion crystal (SkX) is thermodynamically stable. Nevertheless, the 
local edge instability prompts the formation of a helical state which is metastable in this case. 

In a second step, the magnetic field is again increased  to a value $h_f$ at $t = t_f$, and, as a consequence, the helical phase is pushed back towards the edge. For this to happen, it is required that $h_f > h^{\rm cr}_{\rm kink}$ so that the field-polarized state is energetically favoured compared to the CSL state and can exert a positive pressure on the interface. For the simulation in Fig.~\ref{fig7}(e)-(g), we have chosen $h_f = h_i$.
Remarkably, the initial state is however not recovered. An interesting local dynamics governs the fate of the helical fingers as they are pushed towards the edge. When the merons, i.e., the half-skrymions approach the boundary, each of them pulls a second meron out of the edge. Both combine to a skyrmion which gets repelled from the edge by the surface twist of the magnetization. As a result, one obtains a chain of equally spaced skyrmions. Note that $h_f = h_i$ is here located within the regime where the FP phase is thermodynamically stable. The interaction of the receding meron with the spin configuration at the edge thus results in a final state, Fig.~\ref{fig7}(g), that possesses a larger energy than the initial field-polarized state of Fig.~\ref{fig7}(a). 
The spin configuration at the boundary with its surface twist apparently acts as a repulsive potential for the merons hindering them to leave the sample so that the field-polarized thermodynamic groundstate becomes dynamically inaccessible. The systems is thus trapped in a metastable state containing a chain of skyrmions.

The advantage of the protocol \eqref{protocol} is that it is extremely robust.
The final state is, for example, completely independent of the time $t_f$ for which the magnetic field is lowered. 
The precise field values and the size of the damping term is also not important. Another important aspect of this protocol is that during the process, the magnetic configuration is always smooth and singular  intermediate spin configurations, i.e.,  Bloch points are not needed. This is in contrast to processes where skyrmions are created within the bulk\cite{Milde2013,Mochizuki2015}.

\begin{figure}[t]
  \centering
  \begin{minipage}[b]{0.47\textwidth}
    \centering
    \includegraphics[width=0.47\textwidth]{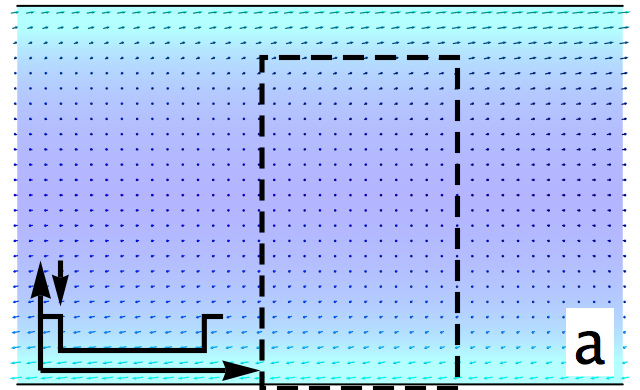}        
    \includegraphics[width=0.47\textwidth]{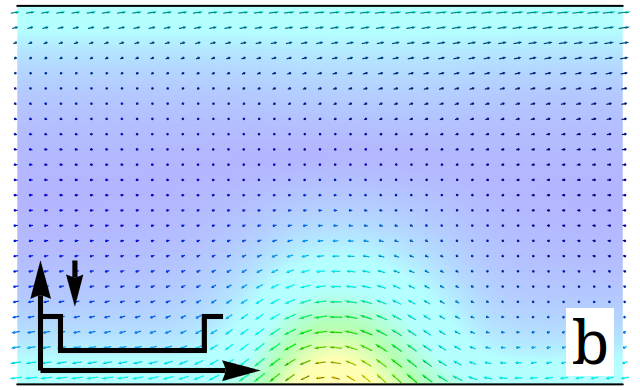}
    \includegraphics[width=0.47\textwidth]{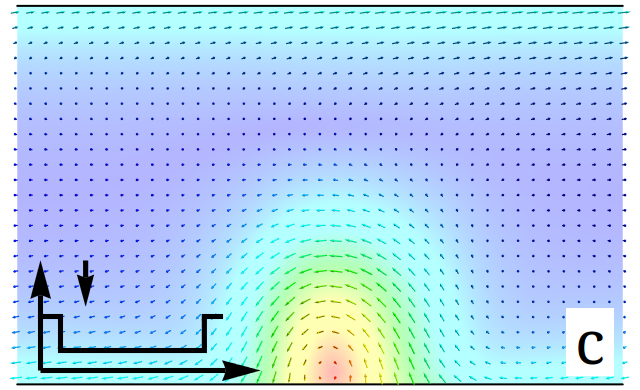}
    \includegraphics[width=0.47\textwidth]{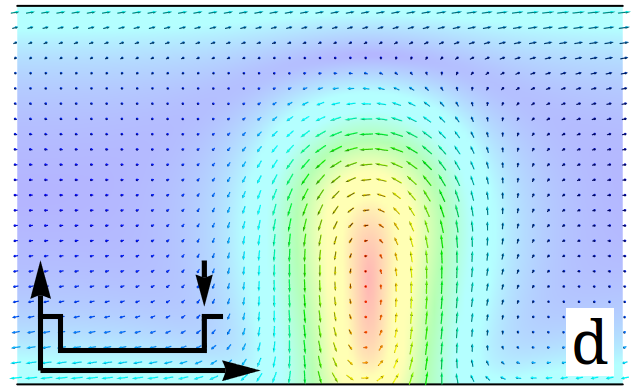}
    \includegraphics[width=0.47\textwidth]{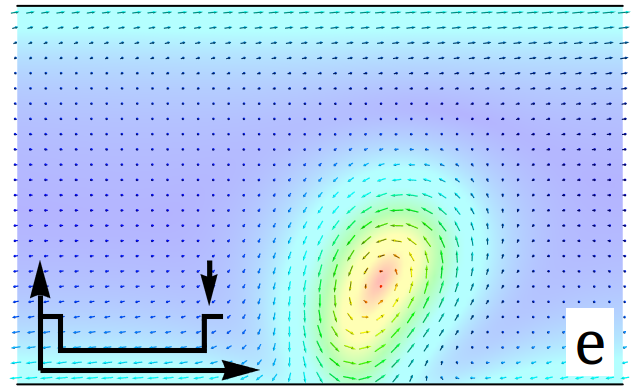}
    \includegraphics[width=0.47\textwidth]{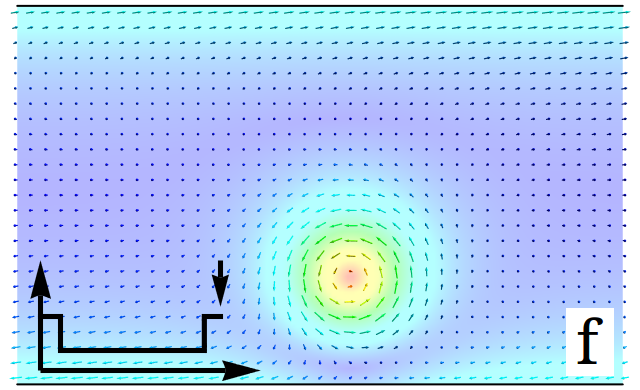}
 \end{minipage}
 \\[2em]
  \includegraphics[width=0.4\textwidth]{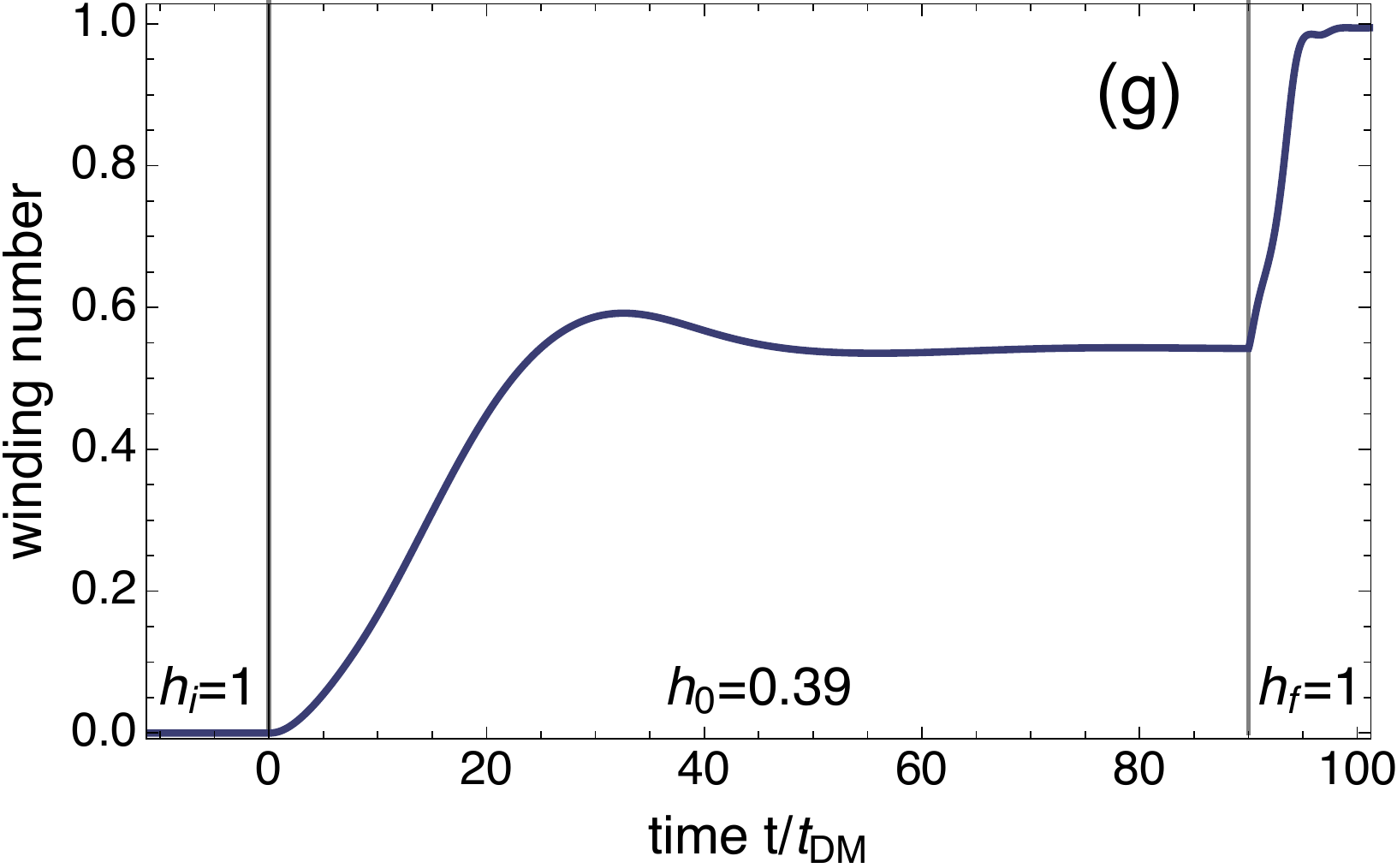}
  \caption{
  Creation of a single skyrmion using the edge instability. 
  We use a similar protocol as in Fig.~\ref{fig7} with $\alpha = 0.4$, $\kappa = 0$, $h_i = 1$, $h_0 = 0.39$, $h_f = 1$ and $\tilde{t}_f = 90$ but this time the magnetic field is only reduced in the finite region marked by the dashed rectangle in panel (a). We have chosen the extent of this region along the edge to be half of the wavelength, $\lambda_x/2=\pi/q_x$, at which the edge magnon softens. The color code in panel (a)-(f) denotes the $z$-component of the magnetization. Panel (g) shows the time evolution of the winding number. }
  \label{fig8}
\end{figure}

The second part of our protocol, the creation of a chain of skyrmions by pushing the helical state towards the edge 
with the help of a magnetic field has  been realized recently for FeGe nanostrips in a beautiful  experiment by Du {\it et al.} in Ref.~[\onlinecite{Du2015}]. In this experiment, the system was first prepared in a helical state. After an increase of the 
magnetic field, helices ultimately reconstruct yielding skyrmions. For nanostrips with a small width, the skyrmions recombined in the center of the strip while for larger width precisely the same process was observed that we discuss above.

The same protocol can also be used to create a single skyrmion by reducing the magnetic field however only within a confined region close to the edge as indicated by the dashed box in Fig.~\ref{fig8}(a). Experimentally, one can use, e.g., a magnetic tip to change locally the magnetic field. Previous micromagnetic simulation by Koshibae {\it et al.} \cite{Koshibae2015} have already demonstrated that  pulses of magnetic field in a small area can trigger the creation of a skyrmion close to the edge. Here, we identify the edge magnon instability as the underlying principle of this phenomenon.
As the field is locally reduced below the edge magnon instability, a meron is protruding into the system as shown in Fig.~\ref{fig8}(b)-(d). As the magnetic field is increased again a single skyrmion forms, see Fig.~\ref{fig8}(e) and (f). In panel (g) we show the time evolution of the total skyrmion winding number. As the field is decreased at time $t=0$, the winding number increases and saturates to a value close to 1/2 reflecting the presence of a single meron. As the field is increased again at $t_f = 90 t_{\rm DM}$, the winding number increases assuming a value close to one when the skyrmion formation is completed.

%----------------------------------------------------------------------------------------
\section{Discussion}
\label{sec:conc}
%----------------------------------------------------------------------------------------

The Dzyaloshinskii-Moriya (DM) interaction imposes boundary conditions on the magnetization\cite{Wilson2013,Rohart2013} which results in a reconstruction of the magnetization profile close to surfaces. For the field-polarized state this just translates to a twist of the magnetization along or perpendicular to the surface normal depending on the type of DM interaction. We have demonstrated that this surface twist can act as an attractive potential on the spin wave excitations leading to magnon modes that are bound to the edge of the sample. The energy of these edge modes as a function of momentum transverse to the edge has been calculated and is shown in Figs.~\ref{fig3} and \ref{fig4}. These bound magnons become soft at a finite momentum thus locally destabilizing the metastable field-polarized phase, and this edge instability triggers the formation of a helical state as shown in Fig.~\ref{fig7}. We have shown by micromagnetic simulations that the process of helix formation via the edge instability is dynamically irreversible allowing for the creation of skyrmions close to the edge. 
%Our results explain the experimental observation of edge-mediated nucleation of skyrmions by Du {\it et al.}\cite{Du2015}.

In the present work, we confined ourselves to a two-dimensional monolayer where the edge instability corresponds to a local instability of the field-polarized state. We now shortly comment on the modifications expected for a magnetic film of finite thickness. In this latter case, the field-polarized state becomes locally unstable upon decreasing the magnetic field with respect to the formation of a conical phase\cite{Wilson2014,Banerjee2014}. Importantly, this instability occurs before the local edge instability discussed in this paper so that our results cannot be directly applied to thin films. However, the conical phase should also exhibit its own edge instability whose analytical description is however more involved due to the coupling of the helical modulation and the surface twist. Nevertheless, we expect the edge instability of the conical phase to possess a similar character also allowing for the creation of skyrmions as discussed above. We thus believe that our results basically explain the mechanism of the edge-mediated nucleation of skyrmions experimentally observed by Du {\it et al.}\cite{Du2015} in thin films of FeGe.

Whereas we focused here on the properties of the field-polarized state, the surface reconstruction is also expected\cite{Rybakov2013,Rybakov2015,Rybakov2016} for the other thermodynamically stable phases, i.e., the skyrmion crystal and the helix. It is an interesting open question how it influences the spin wave spectrum and whether or not bound magnon edge modes also exist in these other phases. At least for the skyrmion crystal phase, magnon edge modes are indeed expected but for a very different reason. Magnons experience an emergent orbital magnetic field when they scatter off a topological skyrmion configuration\cite{Nagaosa2013,Schuette2014,Schroeter2015}. In a magnetic skyrmion crystal this should give rise to a topological magnon band structure characterized by finite Chern numbers and the concomitant topological edge modes\cite{Roldan-Molina2015}.

%----------------------------------------------------------------------------------------
\section{Acknowledgements}
%----------------------------------------------------------------------------------------

J.M. acknowledges helpful discussions with V.~Cros and J.-V.~Kim as well as support from the Deutsche Telekom Stiftung.

\bibliography{edgemagnons}

\end{document}